\documentclass[conference]{IEEEtran}
\IEEEoverridecommandlockouts
\usepackage{cite}
\usepackage{amsmath,amssymb,amsfonts}
\usepackage{algorithmic}
\usepackage{graphicx}
\usepackage{textcomp}
\def\BibTeX{{\rm B\kern-.05em{\sc i\kern-.025em b}\kern-.08em
    T\kern-.1667em\lower.7ex\hbox{E}\kern-.125emX}}
\begin{document}

\title{Analysis of the problem of intervention control in the economy on the basis of solving the problem of tuning\\
}

\author{\IEEEauthorblockN{1\textsuperscript{st} Peter Shnurkov}
\IEEEauthorblockA{\textit{National Research University} \\
\textit{Higher School of Economics}\\
Moscow, Russia \\
pshnurkov@hse.ru}
\and
\IEEEauthorblockN{2\textsuperscript{nd} Daniil Novikov}
\IEEEauthorblockA{\textit{National Research University} \\
\textit{Higher School of Economics}\\
Moscow, Russia \\
even.he@yandex.ru}
}

\maketitle

\begin{abstract}
The paper proposes a new stochastic intervention control model conducted in various commodity and stock markets. The essence of the phenomenon of intervention is described in accordance with current economic theory. A review of papers on intervention research has been made. A general construction of the stochastic intervention model was developed as a Markov process with discrete time, controlled at the time it hits the boundary of a given subset of a set of states. Thus, the problem of optimal control of interventions is reduced to a theoretical problem of control by the specified process or the problem of tuning. A general solution of the tuning problem for a model with discrete time is obtained. It is proved that the optimal control in such a problem is deterministic and is determined by the global maximum point of the function of two discrete variables, for which an explicit analytical representation is obtained. It is noted that the solution of the stochastic tuning problem can be used as a basis for solving control problems of various technical systems in which there is a need to maintain some main parameter in a given set of its values.
\end{abstract}


\section{Introduction}
The article addresses the problem of the development and the analysis of a stochastic model of the intervention phenomenon in economic systems.
As such a model, it is proposed to use a Markov stochastic process with discrete time, that is, a Markov chain. In the scientific literature, papers are known in which Markov processes with discrete time are used to describe the intervention, but all these models are built on the basis of autoregression. A more detailed description of such studies is given in Section III. In this paper, the stochastic intervention model is constructed on the basis of a special type of Markov process, namely, Markov chains with absorption. Another fundamental difference between the proposed model and the known ones is the presence of controls, which are realized after ingression into absorbing states. These controls describe the intervention, that is, external influence on an economic system.

We also note that the proposed stochastic Markov model with discrete time has a universal character and can be used to describe the control of technical systems, in the operation of which it is necessary to maintain a certain basic parameter in given acceptable limits. A more detailed analysis of the possible applications of the developed stochastic model is given in the final part of this work.

Short description of the stochastic model describing the above-mentioned phenomena occurring in technical and economic systems is proposed by P.V. Shnurkov in \cite{SN1}. This publication also presents the results of solving the optimal control problem for this model, which the author has called the tuning problem.

\section{Description of the phenomenon of intervention in the economy}

To design the model we need to describe the process behind it. Hence, let's begin with the definition of the "intervention"\ and other related terms from economics.

Intervention is commonly defined as an economical influence of one entity on actions and matters of another entity through the investment and the deposit of its own funds.

Usually the intervention operations are carried out via central bank and the Treasury by means of massive buying or selling currency, securities and the provision of credit in order to normalize the financial system \cite{SN2}.

As the definition indicated, there are two main types of interventions: one of them has its purpose in selling goods, currencies or securities and the other in buying them. These actions are the essence of the so called "intervention"\ .

Such activities as "intervention purchases"\ and "intervention stocks"\ are taking place in markets for agricultural commodities in Russian Federation and they are also defined by the law. We will refer to these notions following the 14th article of the federal law \cite{SN3}:

The government intervention purchases, stock interventions are carried out with a view to stabilizing the prices in the agricultural market, foods and commodities, as well as to maintaining income-support for agricultural producers.

The government intervention purchases are taking place when prices for agricultural products fall below the estimated prices. The means of conducting such intervention is through purchasing (including exchange trades) agricultural commodities from agricultural producers or implementing pledge operations in relation to the products in question.

Government interventions in the stock markets are carried out through selling the purchased agricultural products (including exchange trades) when the prices on the agricultural products exceed the maximum estimated prices.

Thus, we've described the essence of an economic phenomenon of interventions. In general, it consists in a purposeful influence on the market of a product, currency or securities in form of buying or selling them. When an intervention purchase is conducted, it is related to the time when a market price of a certain product falls below a minimum limit. Intervention stocks will be conducted at times when a market price exceeds the maximum of a predetermined level. In this respect, we should note that the characteristics given above, namely, once price reaches its lower or upper boundary levels, are the main reasons for the intervention. Therefore, the main purpose of the intervention is to reach such a condition of an economic system (in commodities or currency market), where one of its main parameters (price of a product or currency) is found within the permissible limits, which is between its minimum and maximum boundary levels. A mathematical model in intervention research should reflect, above all, the core elements of the real process of intervention.

\section{Review of researches of intervention phenomenon by mathematical methods}

\textbf{Russian research of intervention}. In the Russian scientific literature there are not a lot of works dedicated to the intervention phenomenon research using mathematical models and methods of analysis of economic systems.
Note the research Romanenko I.N. and Evdokimova N.E. \cite{SN4}, \cite{SN5}, \cite{SN6}. These works describe the construction of a mathematical model of intervention purchases and intervention stocks in the Russian grain market.
The basis of this model is the balance principle.

This model is a set of linear relations that include equalities and inequalities between the main dynamic characteristics of the system (regional or All-Russian grain market).
Each relation expresses a certain kind of balance or "partial equilibrium"\ (in the terminology of the authors).
For example, the sale of grain by producers is linearly dependent on the equilibrium price of grain and current stocks of grain producers.
The coefficients of these linear relations are calculated using regression analysis and refined in the analysis of data for 2000-2009.

Analytical representation of market equilibrium price as a function of the volume of grain purchases becomes possible as a result of the analysis of this system of relations. This representation is as follows \cite{SN6}:

$$C(t)=\frac{V(t)(e-b)-d+f+0,001hK_{rub}C_{W}(t)+S(t)}{h+a-g}, \eqno(1)$$

where

$C(t)$ - equilibrium price of grain;

$V(t)$ - current stocks of marketable grain from producers;

$S(t)$ - trade and purchasing balance (positive for purchases);

$K_{rub}(t)$ - ruble-dollar exchange rate;

$C_{W}(t)$ - price of export contracts;

These characteristics depend on the parameter $t$, which has the sense of time,
that is, they can be considered as dynamic.

Constants $a, b, d, e, f, g, h$ included in the expression for the market equilibrium price $C(t)$ are coefficients whose numerical values can be determined on the basis of available information on the grain market.

This relation is the main theoretical result of this research.

According to the authors \cite{SN6}, the use of this relation will solve various problems associated with the effective conduct of intervention purchases and intervention stocks.
In particular, this will allow us to calculate the volumes of interventions aimed at maintaining indicative prices or optimizing the criterion of the efficiency of the functioning of the grain market.
In addition, this will make it possible to assess the effects of a complex regulatory impact on pricing processes and market winnings of all grain market entities, depending on the amount and timing of grain interventions.

Let us give some comments on the above conclusions given in \cite{SN4}, \cite{SN5}, \cite{SN6} concerning the model in question.

\newcounter{numberedCntA}
\begin{enumerate}
\item The presented model is linear and deterministic. Real market processes are random by their nature.
\item Even if we accept the proposed model, there is no explanation why the formula (1), expressing the price of the product, provided that a certain "market equilibrium"\ is fulfilled, allows us to optimize the criterion of the effectiveness of the functioning of the market. Mathematically, the optimization problem is not posed and is not solved here.
\item The conclusion about the possibility of assessing the effects of complex regulatory impact on pricing processes and the market winnings of all grain market entities, depending on the volumes and timing of grain interventions, can only be taken in terms of the ability to numerically evaluate the effect of individual parameters on others in the linear deterministic model.
\setcounter{numberedCntA}{\theenumi}
\end{enumerate}

Despite these features and significant simplifications, the studies carried out in papers \cite{SN4}, \cite{SN5}, \cite{SN6} are significant because they form and analyze one of the first mathematical models describing interventions in the Russian grain market.

A significant research of the Russian grain market using mathematical methods was carried out in \cite{SN7}. The goal of the authors of this work was to find stable modes of operation of the intervention fund, in which the balance between intervention stocks and intervention purchases is maintained in the long run. The general method of D. Forrester's system dynamics is used. A simulation model of the grain market in Russia was developed, based on a system of dynamic linear relations. In the construction of the model, a nonequilibrium approach is used. It is assumed that (a) the demand and supply functions are known; (b) demand and supply at an arbitrarily chosen point in time are generally not balanced, and imbalances are compensated by the dynamics of grain stocks, including the intervention fund. The regression multidimensional dependence of the export size on the gross grain harvest in the preceding months, the foreign trade price of grain and the price of grain in the domestic market was constructed. Estimates of the parameters of this regression model are found. The obtained results are used for qualitative assessments of measures of state regulation of the grain market in Russia. An algorithm for making decisions on interventions in the grain market based on a "floating" price corridor is proposed. Computer experiments on the basis of this algorithm were carried out, according to their results a number of qualitative conclusions were made.

In \cite{SN8} the model of M.Porter's competitive forces for the Russian grain market is constructed. This model is a tool for qualitative analysis. Based on this model, the author makes a number of qualitative conclusions.

There are also special studies devoted to the analysis of interventions in the Russian foreign exchange market.
Among these, we note \cite{SN9}, which analyzed the Bank of Russia's foreign exchange interventions for the period from January 2001 to November 2008.
In this paper, the authors use methods of qualitative economic analysis and classical econometric methods related to the construction of mathematical models.
In this work, data on gross foreign exchange interventions of state funds and the Bank of Russia were collected and systematized.
It is noted that the main parameter, to which the impact of intervention is directed, is the value of the bi-currency basket in rubles, that is, amounts of dollar and euro  with the corresponding weight coefficients.
The ideas used to assess the effectiveness of foreign exchange interventions are characterized.
It has been established that the Bank of Russia foreign exchange interventions do not have a significant impact on the value of the bi-currency basket, and therefore, on the tendency of the ruble exchange rate change in the medium and long term.
But at the same time, these interventions significantly affect the volatility of the value of the bi-currency basket.
The author uses autoregressive models GARCH and GARCH-M, describing volatility and characterizing the impact of interventions on it.
The results of statistical analysis of the quality of the models considered are presented.
We note in addition that the results of the conducted studies of autoregressive models of volatility are given by the authors in a very concise, overview form, without presentation of the original mathematical relations.

\textbf{Foreign research of intervention}. Scientific studies of the phenomenon of intervention in commodity and currency markets are also being conducted in other countries.
At the same time in the foreign scientific literature there are much more publications devoted to the analysis of the phenomenon of intervention than in the Russian.
The bulk of research is focused on two areas.
The first of them is related to the research of interventions conducted by state organizations in the markets of grain crops of the various countries.
The second is devoted to the study of regularities and features of interventions conducted by state banks and other state financial structures in foreign exchange markets.
Let's give a brief description of some of them, in which mathematical methods are used to some extent.
Let's start with the research of interventions in grain markets.

Work \cite{SN10} is devoted to the research of the grain market in China. To describe the grain market, a multidimensional autoregressive linear model is constructed.
This model includes five main characteristics that describe the market.
The model is discrete in time, the values of the main indicators are determined annually on the interval from 1986 to 1998.
The model takes into account the forms of state impact on the market
(quotas for government purchases of grain, set by the government).
The import of cereals also influences the market.
The authors assume that the values of the main indicators at the current time $t$ can depend on the values of the other indicators at the moments $t$ and $t-1$. The paper describes obtaining numerical values of estimates of the coefficients of the autoregressive model.
The importance of influence of some indicators on others is investigated.

This autoregressive model of the grain market in China is used to simulate the possible effects of the state on the state of the market.
The author's calculations show the consistency of the results obtained in the modeling and the data on the actual processes that took place in the Chinese grain market in the specified period of time.

We also note that the bibliography of scientific research of the grain market in China is given in the work \cite{SN10}.

A thorough and extensive study of the various forms of state influence on the rice market in India is conducted in \cite{SN11}.
The author considers two main forms of such impact: state purchases of raw (unpurified) rice grain directly from producers (farms) at guaranteed minimum government support prices (MSP) and the purchase of processed (purified) rice from primary grain processors (mills) using preferential system of taxation.
Thus, direct and indirect forms of support of farms are provided and production is stimulated.

The paper uses a multidimensional stochastic market model that describes the dependence of the relative capital growth of the main participants (producers of goods) on the volume of investments (purchases) of buyers.
This dependence is expressed in an explicit linear analytical form and has a dynamic character, that is, it takes into account the dependence on the transaction number (time point).
We should especially note that in the model all buyers of grain are divided into two groups: large and small, which is taken into account in the nature of the functional dependence.
In this paper, it is assumed that the distribution density of procurement volumes can be approximated by means of Hermite series of order 2.
On the basis of the statistical material, estimates of the unknown parameters of the model are constructed.
After evaluating unknown parameters and specifying the model, the author of the work uses it to simulate real processes in the rice market.
Using modeling, estimates of the incomes of grain producers are made and the dependence of these incomes on various factors is investigated: minimum government procurement prices, levels of taxes on product prices of primary processors, and the level of competition in the market determined by the number of participants in transactions.
The authors of the work calculate the observed prices and the corresponding estimates of the producers' incomes.
Based on the methods of statistical analysis, the asymptotic distribution of the optimal purchase price estimate is determined.
Thus, it becomes possible to determine the estimate for the optimal procurement state price and compare the producers' incomes in real conditions using the optimal state procurement prices.

The paper \cite{SN12} investigates some of the special effects associated with interventions in the market for genetically unmodified soybeans at the Tokyo Stock Exchange.
In particular, the effect of the receipt of information on the change in the duration of the futures contracts on the highest price of this type of grain is investigated.
The autoregressive integrated moving average model (ARIMA) from the theory of time series has been used as a mathematical model to describe the intervention impact on the price.
As a result of the analysis on specific data, it is established that incoming information leads to an increase in the price, but this effect manifests itself with a delay equal to four months.
This means that the market in question is not effective, because in theory, an effective market should immediately respond to incoming information.
In addition to this result, this work is interesting because in the model under consideration, the role of direct impact, that is, intervention, is not the volume of real purchases or sales, but information about changing the "rules of the game" in the relevant market.

In \cite{SN13} a mathematical model of the housing market is proposed, based on the idea of general equilibrium. In the role of external influences (interventions) are tax benefits associated with income from employment, income from capital and rental income. The influence of these interventions on the welfare of buyers is investigated. In particular, home buyers can perform market analysis and determine the most preferable forms of action in the market: the acquisition of housing in the property, or rent. In addition, homeowners are given the opportunity to assess the effectiveness of additional investments in housing and subsequent rental housing.

Note also the paper \cite{SN14}, in which an extensive review of econometric models and methods used to describe the impact of the state in various areas of the economy is made.

Summarizing the review, we will give a general description of scientific research on the phenomenon of intervention in the economy, conducted using mathematical models and methods.

Let us note first of all that the problem of investigating interventions in economic systems is important and urgent.
It attracts the attention of specialists in various countries.

In the modern scientific literature there are studies in which crops mathematical models of interventions in the markets of grain.
Such models are inherently stochastic, given the random nature of the factors operating in a free competitive market.
They are regressive and autoregressive relations describing the impact of various factors, including the extent of interventions, on some of the key indicators that characterize the market system under consideration.
The unknown parameters included in these relations are estimated from the results of observations, that is, according to the statistical information at the disposal of the authors.
After the creation of such a model, it becomes possible to assess the impact of interventions on the most important basic economic indicators of the system.
There are other related studies in \cite{SN10}, \cite{SN11}, \cite{SN12}, \cite{SN7}. There are also works based on the deterministic linear model of the grain market (\cite{SN4}, \cite{SN5}, \cite{SN6}).

\ \ \ \ Investigations of the phenomenon of intervention in currency markets using mathematical methods are rarely conducted \cite{SN15}. In our opinion, this is due to the objectively existing complexity of this phenomenon, as well as the fact that decisions on such interventions are made at the government level and depend not only on economic but also on political factors that are difficult to describe.

\section{General structure and basic features of the stochastic model
interventions in the economy}

In this section, a general concept of a new stochastic model will be proposed, intended to describe the phenomenon of intervention. This concept is based on the general concept of intervention in the economic system formulated in Section II, as well as on a qualitative analysis of this phenomenon.

In the framework of this study, the economic system will be understood as a certain commodity or financial market in which random factors objectively act. In this market, some basic parameter is formed that changes over time. This parameter is a stochastic process that will be considered a mathematical model of the functioning of the system under research. The state of this process at an arbitrary point in time will characterize the state of the system.
In the commodity market system (for example, the grain market) this parameter is the current unit price of the relevant product.
In the grain market system, this parameter is the product price. This parameter may be used as the present value of the bi-currency basket in the financial market system.

Evolution of this process
consists of two main stages that go one after another and
forming a repeating cycle. At the first stage, the price is formed without
external influence, according to the internal rules of this market
system. The set of possible states of the process includes the specified subset of admissible states. External influence is not used during the process stay in this admissible subset. At moments when the process (the main parameter) goes to the border or beyond the subset of admissible states, an external influence is made. This is the intervention. Conducting such an impact forms the second stage of the process evolution. The purpose of this influence
is the return of the process value to one of the state from admissible subset. This external influence is the control in this
probability model. The problem of optimal control is
the choice of control characteristics that give
an extremum to some indicator of control quality. This indicator should have an economic background and reflect the efficiency of the economic system.

We take the following important assumption related to the nature of the proposed stochastic model. This assumption is that the process describing the functioning of this system has a Markov property. As is known, the Markov property consists in the fact that after entering a certain fixed state, the further evolution of the process occurs independently of the past, and depends only on the specified state. This property is characteristic of many economic and technical systems and related stochastic processes. The presence of this property will allow to use the results of a deeply developed theory of Markov processes.

Illustration of the concept described above - Fig. 1,
which shows the possible trajectory of the main process and
possible control effect.

\begin{figure}[htbp]
\centerline{\includegraphics[width=0.5\textwidth]{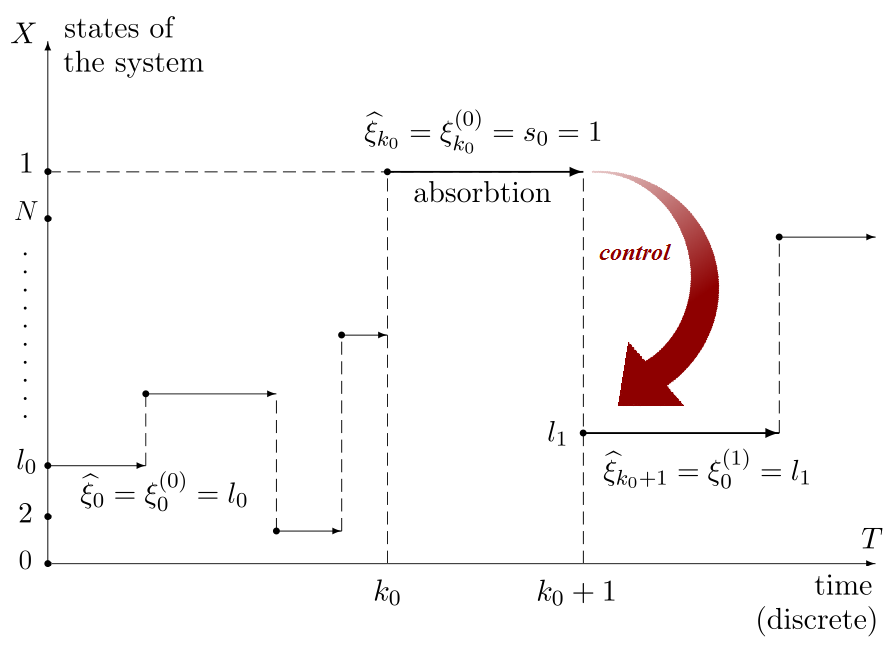}}
\caption{A possible trajectory of the stochastic process $\left\{ {\widehat{\xi}
}_{k}\right\} $, which is a stochastic model of the behavior of the main
parameter.}
\label{fig}
\end{figure}

Let's give some comments to Figure 1, explaining the aspects of the above stochastic model.
However, we will first accept certain conventions on the designation of states.

Under the natural order of numbering, the boundary states for the set in question are the states $\{0\}$ and $\{N\}$. In this model, the boundary states are considered invalid, and the internal states $\{1,2,\dots,N-1\}$ - are admissible. However, in order to build a subsequent theory, it will be necessary to immediately re-designate the states, abandoning the natural numbering. Further in this investigation we will use the classical theory of absorbing Markov chains \cite{SN16}.
For convenience in applying this theory, we rewrite the states in the original set $\lbrace 0,1,2,\ldots ,N\rbrace $  as follows: the state $\lbrace 0\rbrace $ we leave the previous notation $\lbrace
0\rbrace $, the state $\lbrace N\rbrace $ will be denoted by the symbol $\lbrace
1\rbrace $, and assign the new notation to the remaining states $\lbrace 1,2,\ldots ,N-1\rbrace $ as $\lbrace 2,3,\ldots ,N\rbrace $. Thus, in the set of states $X=\lbrace
0,1,\ldots ,N\rbrace $ the states $\lbrace 0\rbrace $ and $\lbrace
1\rbrace $ will be boundary and absorbing states, and the states $
\lbrace 2,3,\ldots ,N\rbrace $ - internal, allowable, non-returnable.

At the initial moment $t_{0}=0$ the process starts from an admissible state $l_0\in\lbrace 2,3,\ldots ,N\rbrace $.
The evolution of the process is described with
absorbing Markov chain $\left\{ \xi _{k}^{(0)}\right\}_{k=0}^{\infty} $, in
which boundary states are absorbing, and internal
admissible states.
As is known from Markovian processes theory
, at some moment $k_{0}$
the process falls into one of the absorbing states; in this example $\xi
_{k_0}^{(0)}=1$. After that, an external influence
(control), as a result of which the process is translated into some
the internal admissible state $l_1\in\lbrace 2,3,\ldots ,N-\rbrace $ with probability $\alpha _{l}^{(1)}$; $\sum\limits_{l=2}^{N}{\alpha
_{l}^{(1)}}=1$.
A similar external influence produced when the process
is absorbed in state 0, i.e. if $\xi _{k_0}^{(0)}=0$,
 is described by the discrete probability distribution $\left( \alpha
_{l}^{(0)},l=2,3,\ldots ,N\right) $.
After each influence, irrespective of the past the process begins to evolve from the state $l_1$ along the trajectory of the new absorbing Markov chain $\left\{ \xi _{k}^{(1)}\right\}_{k=0}^{\infty} $, whose probabilistic characteristics coincide with the characteristics of the chain $\left\{ \xi _{k}^{(0)}\right\}_{k=0}^{\infty} $.
At the moment of falling into one of the boundary absorbing states, an external control influence is again performed, which is described with the discrete probability distributions $\left( \alpha _{l}^{(0)},l=2,3,\ldots ,N\right) $, $\left( \alpha _{l}^{(1)},l=2,3,\ldots ,N\right) $.

\section{Formal construction of a stochastic model in the form of a Markov process with discrete time}

We now turn to the formal construction of a stochastic Markov model with discrete time and a finite set of states $X=\{0,1,\dots,N\}$, describing the functioning of a system with periodic external influences.

We note first that all stochastic objects introduced in the future are assumed to be given on the same initial probability space $\left(\Omega,\textit{A},\textbf{\textit{P}}\right)$. This space formalizes a random experiment conducted in objective reality with the system in question. The concept of probability space and its properties are described in detail, for example, in research \cite{SN17}, \cite{SN18}, \cite{SN19}.

Suppose that a sequence of independent Markov chains $\left\{\xi_k^{(n)}\right\}_{k=0}^{\infty},~n=0,1,2,\dots$ is given. We note in particular that these chains are uncontrollable and describe the evolution of the system under consideration over time periods between successive external influences.

Following the general concept of the proposed stochastic model described in Section IV, the evolution of a discrete-time Markov process, which will play the role of the basis of this model, can be described as follows.
Suppose that some Markov chain $\left\{ \xi _{k}^{(n)} \right\}_{k=0}^{\infty}$ with fixed number $n$ begins to evolve in one of the admissible states
$\lbrace 2,3,\ldots ,N\rbrace $. After a finite time after the beginning of the evolution, the process $\left\{ \xi _{k}^{(n)} \right\}_{k=0}^{\infty}$ with probability $1$ falls in one of the boundary states $\lbrace 0\rbrace $ or $\lbrace 1\rbrace $.
After the process $\left\{ \xi _{k}^{(n)} \right\}_{k=0}^{\infty}$ is absorbed, the model is subjected to an external influence, which consists in the transition from the absorbing state to one of admissible (internal) states.
Note that such an assumption is fundamentally important for building a model. His analytical expressions and probabilistic content will be explained later.
After the transition to the admissible state $l\in\lbrace 2,3,\ldots
,N\rbrace $ the evolution of the system will continue irrespective of the past and described by the Markov chain $\left\{ \xi _{k}^{(n+1)} \right\}_{k=0}^{\infty}$ whose probabilistic characteristics coincide with the characteristics of the process $\left\{ \xi _{k}^{(n)} \right\}_{k=0}^{\infty}$. The further evolution of the process describes analogously.

Let $\nu^{(n)}$ denote the random time from the beginning of the evolution of the process $\left\{ \xi _{k}^{(n)} \right\}_{k=0}^{\infty}$ in one of the internal states of $\{2,3,\dots,N\}$ to the moment of absorption, $n=0,1,2,\dots$.
It follows from the properties of the absorbing Markov chains $\left\{\xi_k^{(n)}\right\}_{k=0}^{\infty},~n=0,1,2,\dots$ that the random variables $\left\{\nu^{(n)},n=0,1,2,\dots\right\}$ are independent and with a probability equal to one, take only finite values. The distributions of these quantities are the same and depend on the initial value of the corresponding Markov chains
$\left\{ \xi _{k}^{(n)} \right\}_{k=0}^{\infty}$, $n=0,1,2,\dots$. We introduce a sequence of random variables $\widehat{\nu}^{(n)},~0,1,2,\dots$, defined by the relations $$
\widehat{\nu}^{(0)}=\nu^{(0)}~,
$$
$$
\widehat{\nu}^{(n)} = \sum\limits_{i=0}^{n}\nu^{(i)}+n~,
$$
$$
n=1,2,\dots~.
$$

Now we define a random process with discrete time $\left\{ \widehat{\xi} _{k}\right\}_{k=0}^{\infty}$, which we will call the main one in the following. We fix some initial state of the process $\left\{\xi_{k}^{(0)} \right\}_{k=0}^{\infty}: \xi_0^{(0)}=l_0\in\{2,3,\dots,N\}$ and assume that the initial state of the process
$\left\{ \widehat{\xi}_{k} \right\}_{k=0}^{\infty}$ coincides with the indicated state: $\widehat{\xi}_0=\xi_0^{(0)}=l_0$. The further evolution of the process $\left\{ \widehat{\xi}_{k} \right\}_{k=0}^{\infty}$ over a period of time before the first entry into one of the boundary states is determined as follows
$$
\widehat{\xi}_k=\xi_k^{(0)}, k=0,1,2,\dots,~ \widehat{\nu}^{(0)}. \eqno(1)
$$

Suppose that the condition $\widehat{\nu}^{(0)}=\nu^{(0)}=k_0$ is satisfied. At the time of $k=k_0$ the process $\left\{ \widehat{\xi}_{k} \right\}_{k=0}^{\infty}$ is in one of the boundary states $s_0\in \{0,1\}$. The behavior of the process $\left\{ \widehat{\xi}_{k} \right\}_{k=0}^{\infty}$ after getting into this state does not depend on the past, that is, on its behavior up to the moment $k_0$, and is determined by the ratio.
\small$$
P\left(\widehat{\xi}_{k_0+1}=l_1|\widehat{\xi}_{0}=l_0,\widehat{\xi}_{1}=i_1,\dots,\widehat{\xi}_{k_0-1}=i_{k_0-1},\widehat{\xi}_{k_0}=s_0\right)=
$$ \normalsize
$$
P\left(\widehat{\xi}_{k_0+1}=l_1|\widehat{\xi}_{k_0}=s_0\right)=\alpha_{l_1}^{(s_0)}, \eqno(2)
$$
$$
l_1\in\{2,3,\dots,N\},~s_0\in\{0,1\},
$$
where $i_1, i_2, \dots, i_{k_0-1}\in\{2,3,\dots,N\}$ -- arbitrary states that form the trajectory of the process $\left\{\widehat{\xi}_k\right\}_{k=0}^{\infty}$ on the time period $\{1,2,\dots,k_0-1\}$.

Provided that $\widehat{\xi}_{k_0+1}=l_1$, we assume $\xi_0^{(1)}=\widehat{\xi}_{k_0+1}=l_1$ and, further,
$$
\widehat{\xi}_{k_0+k+1}=\xi_k^{(1)},~k=1,2,\dots,\nu^{(1)} \eqno(3)
$$

Suppose that for some value $n$, $n=0,1,2,\dots$ the moment of the $n$-th hit of the main process $\left\{\widehat{\xi}_k\right\}_{k=0}^{\infty}$ to one of the boundary states is fixed: $\widehat{\nu}^{(n)}=k_n$, and $\widehat{\xi}_{k_n}=s_n,~s_n\in\{0,1\}$.
Then the further behavior of the $\left\{ \widehat{\xi}_{k} \right\}_{k=0}^{\infty}$ process does not depend on the past and is determined by the ratio
\small
$$
P\left(\widehat{\xi}_{k_n+1}=l_{n+1}|\widehat{\xi}_{0}=l_0,\widehat{\xi}_{1}=i_1,\dots,\widehat{\xi}_{k_n-1}=i_{k_n-1},\widehat{\xi}_{k_n}=s_n\right) =
$$
\normalsize
$$
P\left(\widehat{\xi}_{k_n+1}=l_{n+1}|\widehat{\xi}_{k_n}=s_n\right) = \alpha_{l_{n+1}}^{(s_n)}, \eqno(4)
$$
$$
l_{n+1}\in\{2,3,\dots,N\},~s_n\in\{0,1\},
$$
where $i_1,i_2,\dots,i_{k_n-1}$ -- arbitrary states that form the trajectory of the process $\left\{\widehat{\xi}_k\right\}_{k=0}^{\infty}$ on the time period $\{1,2,\dots,k_n-1\}$.

Provided that $\widehat{\xi}_{k_n+1}=l_{n+1}$,  we assume
$$
\xi_0^{(n+1)}=\widehat{\xi}_{k_n+1}=l_{n+1},
$$
and the subsequent trajectory of the process $\left\{\widehat{\xi}_k\right\}_{k=0}^{\infty}$ process is determined by the relation
$$
\widehat{\xi}_{k_n+k+1}=\xi_k^{(n+1)},~k=1,2,\dots,\nu^{(n+1)} \eqno(5)
$$

Relations (1) - (5) completely describe the behavior of the main stochastic process $\left\{\widehat{\xi}_k\right\}_{k=0}^{\infty}$.

A random sequence $\left\{\widehat{\xi}_k\right\}_{k=0}^{\infty}$ is a Markov chain. This follows from the Markov property of the sequences $\left\{\xi^{(n)}_k\right\}_{k=0}^{\infty},~n=0,1,2,\dots$ and the law of transitions from boundary to internal states, which are determined by relations (2), (4). The random variables $\widehat{\nu}^{(n)},~n=0,1,2,\dots$ are the points in time at which the process $\left\{\widehat{\xi}_k\right\}_{k=0}^{\infty}$ reaches the boundary values of the set of states, that is, one of the $\{0,1\}$ states. After reaching an arbitrary boundary state, a control is made, which consists in transferring the process $\left\{\widehat{\xi}_k\right\}_{k=0}^{\infty}$ to one of the internal states. This transferring is described by a probability distribution $\alpha^{(0)}=\left(\alpha_l^{(0)},~l=2,3,\dots,N\right)$, $\alpha^{(1)}=\left(\alpha_l^{(1)},~l=2,3,\dots,N\right)$.
We say that a given pair of probability distributions $\left(\alpha^{(0)},\alpha^{(1)}\right)$ forms a strategy for controlling the main random process $\left\{\widehat{\xi}_k\right\}_{k=0}^{\infty}$.
At the same time, the process $\left\{\widehat{\xi}_k\right\}_{k=0}^{\infty}$ is uncontrollable over time periods between hits to the boundary states.

The mathematical problem of optimal control in this stochastic model is to find a pair of probability distributions
$\alpha^{(0)}=\left\{\alpha_l^{(0)},~l=2,3,\dots,N\right\}$, $\alpha^{(1)}=\left\{\alpha_l^{(1)},~l=2,3,\dots,N\right\}$, which deliver an extremum to some stationary cost indicator of efficiency. The definition of such an indicator and the formal formulation of the optimal control problem will be given later.

We introduce another auxiliary stochastic object, namely, the random sequence $\left\{\zeta_n\right\}_{n=0}^{\infty}$, associated with the main process $\left\{\widehat{\xi}_k\right\}_{k=0}^{\infty}$. We will assume that the process $\left\{\zeta_n\right\}_{n=0}^{\infty}$ is determined by the ratio $\zeta_n=\widehat{\xi}_{\widehat{\nu}^{(n)}},~n=0,1,2,\dots$.

Thus, the values of the random sequence $\left\{\zeta_n\right\}_{n=0}^{\infty}$ coincide with the values of the main process at the moments when the latter falls into the boundary states. Obviously, the elements of the random sequence $\left\{\zeta_n\right\}_{n=0}^{\infty}$ take values in the set $Z=\{0,1\}$.

Note that the introduced sequence of random variables $\left\{\zeta_n\right\}_{n=0}^{\infty}$ forms a Markov chain. Indeed, we fix a random time point $\widehat{\nu}^{(n)}=k_n$ and the state of the process $\zeta_n = \widehat{\xi}_{k_n}=s_n\in\{0,1\}$.
Then the further evolution of the process $\left\{\zeta_n\right\}_{n=0}^{\infty}$ will be determined by the evolution of the main process $\left\{\widehat{\xi}_k\right\}_{k=0}^{\infty}$ after the time point $k_n$. In turn, the evolution of the process $\left\{\widehat{\xi}_k\right\}_{k=0}^{\infty}$ immediately after the $k_n$ moment will be determined by the probability distribution $\alpha^{(s_n)}=\left(\alpha_l^{(s_n)},~l=2,3,\dots,N\right)$ and the Markov chain transition probability matrix $\left\{\xi_k^{(n+1)}\right\}_{k=0}^{\infty}$.

Let us explain the last remark in more detail. Under the above conditions, the initial state of the chain $\left\{\xi_k^{(n+1)}\right\}_{k=0}^{\infty}$ is determined by the distribution of $\alpha^{(s_n)}=\left(\alpha_l^{(s_n)},~l=2,3,\dots,N\right)$. Further, at a fixed initial state $\xi_0^{(n+1)} = l_{n+1}\in\{2,3,\dots,N\}$ the probability that the chain $\left\{\xi_k^{(n+1)}\right\}_{k=0}^{\infty}$ enters one of the absorbing states $s_{n+1}\in\{0,1\}$ is determined.

This characteristic is called the absorption probability and is expressed in terms of the elements of the transition matrix of the Markov chain $\left\{\xi_k^{(n+1)}\right\}_{k=0}^{\infty}$, the corresponding formula will be given in the next section. Thus, under the conditions $\widehat{\nu}^{(n)}=k_n$ and $\zeta_n = \widehat{\xi}_{k_n} = s_n$ the value of the chain $\left\{\zeta_n\right\}_{n=0}^{\infty}$ at the next time instant $\zeta_{n+1} = s_{n+1}$ does not depend on the past and is determined by the given probability characteristics.

In accordance with the terminology adopted in the theory of Markov processes, the introduced Markov chain $\left\{\zeta_n\right\}_{n=0}^{\infty}$ can be called a chain embedded in the main random process $\left\{\widehat{\xi_k}\right\}_{k=0}^{\infty}$. A nested Markov chain $\left\{\zeta_n\right\}_{n=0}^{\infty}$ will play an important role in analyzing the properties of the constructed stochastic model and determining the necessary probability characteristics.

\section{Main characteristics of the stochastic model}

In this section, theoretical results for absorbing Markov chains will be used \cite{SN16}.

We assume that for Markov chains $\left\{ \xi _{k}^{(n)} \right\}_{k=0}^{\infty}$, $n=0,1,\dots$
the following matrix probabilistic characteristics are given:

$\textbf{P}_{00}$ - matrix of transition probabilities within the set of admissible states $\lbrace 2,3,\ldots ,N\rbrace $, has dimension $(N-1)\times (N-1)$;

$\textbf{P}_{01}$ - matrix of transition probabilities from admissible states $
\lbrace 2,3,\ldots ,N\rbrace $ to absorbing states $\lbrace
0,1\rbrace $ in one step of the chain, has dimension $(N-1)\times 2$;

$\textbf{P}_{10}$ - is the matrix of transition probabilities from absorbing states $
\lbrace 0,1\rbrace $ to the admissible states $\lbrace 2,3,\ldots
,N\rbrace $. This matrix is zero, has dimension $2\times (N-1)$;

$\textbf{P}_{11}$ - matrix of transition probabilities from absorbing states $\lbrace 0,1\rbrace $. This matrix is single, has dimension $
2\times 2$.

Then the transition probability matrix of the Markov chain $\left\{ \xi _{k}^{(n)} \right\}_{k=0}^{\infty}$ with an arbitrary number $n$ has the following cellular structure

 $\textbf{P}=\begin{pmatrix}
\textbf{P}_{11} & \textbf{P}_{10} & \\
\textbf{P}_{01} & \textbf{P}_{00} & \\
\end{pmatrix}$

Suppose that the following cost characteristics of the model are given.

Denote as $c_{l}$ the income for a single stay of the main process $\left\{\widehat{\xi}_k\right\}_{k=0}^{\infty}$ in the state $l\in\lbrace 2,3,\ldots ,N\rbrace $ in the period of free evolution (without external influences).
Let $\bar{c}=(c_{l},l\in\lbrace 2,3,\ldots ,N\rbrace )^{T}$ be the column vector of these revenues.

Let denote as $d_{i}^{(s)}$ the value of the costs associated with transferring the main process from the boundary state $s$ to the internal state $i, s\in\lbrace 0,1\rbrace , i\in\lbrace
2,3,\ldots ,N\rbrace $. These costs characterize the price of external influence, which determines the transfer of the process. In accordance with their economic content, these values are negative.

We obtain representations for some additional probabilistic and cost characteristics of the model in question, based on the theory of absorbing Markov chains \cite{SN16}.

Let $b_{i0},b_{i1}$ be the probabilities of absorption of the Markov chain $\left\{ \xi _{k}^{(n)} \right\}_{k=0}^{\infty},~n=
0,1,2,\ldots $ in states $\lbrace 0\rbrace $ and $\lbrace
1\rbrace $ respectively, provided that at the initial time
this process is in the $i$ state; $ \xi _{0}^{(n)}=i,
i\in\lbrace 2,3,\ldots ,N\rbrace $; note that $
b_{i1}=1-b_{i0}, i\in\lbrace 2,3,\ldots ,N\rbrace $.

Further, $r_{i}$ is the expectation of the income associated with the behavior of the Markov chain $\left\{ \xi _{k}^{(n)} \right\}_{k=0}^{\infty},~n = 0,1,2,\ldots$  for a period of time before absorption, provided that at the initial moment of time this process is in the state $i$; $ \xi _{0}^{(n)}=i, i\in\lbrace 2,3,\ldots
,N\rbrace $.
In this model, it is assumed that the income associated with the behavior of the Markov chains $\left\{\xi_k^{(n)}\right\}_{k=0}^{\infty}$, $n=0,1,2,\dots $, defined by the parameters $\overline{c}=\left(c_l,~l\in\{2,3,\dots,N\}\right)^{\mathrm{T}}$, given above, generate incomes on the corresponding trajectories of the main process $\left\{\widehat{\xi}_k\right\}_{k=0}^{\infty}$.

Then the absorption probability matrix $\textbf{B}=(b_{i0}, b_{i1}, i\in\lbrace 2,3,\ldots ,N\rbrace )^{\mathrm{T}}$ is defined by the formula
 $\textbf{B}=(\textbf{I}-\textbf{P}_{00})^{-1}\textbf{P}_{01}$,

where $\textbf{I}$ - is unit matrix of the corresponding dimension.

The vector $\bar{r}=(r_{i},i\in\lbrace 2,3,\ldots ,N\rbrace )^{T}$ can be expressed as follows
$\bar{r}=(\textbf{I}-\textbf{P}_{00})^{-1}\bar{c}$.

Thus, to obtain subsequent results on solving the optimal control problem in the stochastic model under consideration, it is necessary to specify the transition probability matrix $\textbf{P}$, the income vector characterizing the evolution of the main process over time periods without external influences $\bar{c}$, and the set of $d_i^{(s)}$, $i\in\{2,3,\dots,N\}$, $s\in\{0,1\}$, values characterizing the costs of external influences or main process controls. The remaining necessary probabilistic and cost characteristics are determined on the basis of the above analytical formulas.

\section{On the analytical presentation of the stationary cost indicator of management efficiency}

We now turn to investigation of the control problem in a stochastic Markov model with discrete time and periodic outputs to the state-set boundary described in Section IV of this paper. For this purpose, we introduce the concept of a cost additive functional associated with a Markov chain. This functional describes a random income or profit arising from the evolution of the corresponding economic system. Also functionals are known in the scientific literature (for example \cite{SN17}, chapter 8).

Let there be given a Markov chain $\left\{ \theta _{k} \right\}_{k=0}^{\infty} $ with a discrete set of states $Z=\lbrace 0,1,2,\ldots \rbrace $. The set of states of a given chain can be finite or countable.

Let define a function $D:Z\to R$, which can also be defined as the set of its values  $(\rho_{i}=D(i), i\in Z)$.
We will interpret $\rho_{i}$ as revenue (positive or
negative) obtained by one stay of the process $\left\{ \theta _{k} \right\}_{k=0}^{\infty} $ in state $i,i\in Z$. Consider a random sequence

$$\gamma _{n}=\sum_{k=0}^{n}{D}(\theta _{k}),n\in\lbrace 0,1,\ldots \rbrace$$

\textbf{Definition 1}. A random sequence (discrete-time process) $\left\{ \gamma _{n} \right\}_{n=0}^{\infty} $ will be called the cost additive functional associated with the Markov chain $\left\{ \theta _{k} \right\}_{k=0}^{\infty} $.

In the course of the further presentation we will use various concepts and properties from the theory of Markov chains. A detailed presentation of this theory can be found in the following fundamental editions: \cite{SN16}, \cite{SN18}, \cite{SN19}, \cite{SN20}.

We formulate a statement characterizing the behavior of the cost additive functional $\left\{ \gamma _{n} \right\}_{n=0}^{\infty} $ for a long evolution of the process $\left\{ \theta _{k} \right\}_{k=0}^{\infty} $ . Statements of this kind are usually called ergodic theorems. Let us quote this statement, following \cite{SN17}, chapter 8.

\textbf{Theorem}.

Let the Markov chain $\left\{ \theta _{k} \right\}_{k=0}^{\infty} $be irreducible, recurrent, and positive. Suppose also that the following condition holds

$$\sum_{k=0}^{\infty }{\vert {\rho}_{k}\vert \pi _{k}<\infty },$$

where $\pi =(\pi _{k} ,k\in Z)$ is the stationary distribution of the Markov chain
$\left\{ \theta _{k} \right\}_{k=0}^{\infty} $. Then for any initial state $
i_{0}\in Z$ the following relation holds.

 $$I = \lim _{n\to \infty }\frac{1}{n}E[\gamma _{n}|\theta
_{0}=i_{0}]=\sum_{i\in Z}{{\rho}_{i}}\pi _{i} \eqno(6)$$

It is natural to call the value on the right-hand side of relation (6) the average stationary specific income associated with the Markov chain $\left\{ \theta _{k} \right\}_{k=0}^{\infty} $.

Returning to the study of the introduced stochastic model with discrete time and controls at the time when the main process $\left\{\widehat{\xi}_k\right\}_{k=0}^{\infty}$ reaches the boundary of a given subset of the set of states, we will consider the Markov chain $\left\{\zeta_n\right\}_{n=0}^{\infty}$ built in Section V embedded in the main process as the Markov chain $\left\{\theta_k\right\}_{k=0}^{\infty}$. The cost characteristics of the model, defined in Section VI, will define the additive cost functional associated with the main process $\left\{\widehat{\xi}_k\right\}_{k=0}^{\infty}$ and the nested Markov chain $\left\{\zeta_n\right\}_{n=0}^{\infty}$. According to its economic content, this functionality will be a random profit accumulated over a certain period of time. Under some conditions, which will be formulated below, one can apply the reduced ergodic theorem. Then the value $I$, determined by the relation (6), will have the meaning of the average specific profit.

Following the classical papers on the theory of control of Markovian and semi-Markov stochastic processes (\cite{SN22}, \cite{SN23}), we consider the value $I$ as an indicator of control effectiveness in the model under consideration.
Note that similar performance indicators are considered in modern studies on the theory of stochastic control: \cite{SN24}, \cite{SN25}.

In the future it will be proved that the stationary cost indicator (6) can be represented explicitly, through the initial probabilistic and cost characteristics of the model, defined in Section VI. We note that this indicator depends on the discrete probability distributions $
\alpha ^{(0)}=\left(\alpha _{i}^{(0)} , i\in\lbrace 2,3,\ldots ,N\rbrace \right),~~ \alpha
^{(1)}=\left(\alpha _{j}^{(1)} , j\in\lbrace 2,3,\ldots ,N\rbrace \right)$ , that specify basic stochastic process control strategy $\left\{\widehat{\xi}_k\right\}_{k=0}^{\infty}$ and nested Markov chain $\left\{\zeta_n\right\}_{n=0}^{\infty}$.
Thus, the optimal control problem in this model can be formulated as the extremal problem
$$
I=I\left(\alpha ^{(0)},\alpha ^{(1)}\right)\to extr , \left(\alpha ^{(0)},\alpha
^{(1)}\right)\in\Gamma _{d} \eqno(7)
$$
where $\Gamma _{d}$ is the set of pairs of vectors of discrete probability distributions defined on a finite set of admissible controls $
U=\lbrace 2,3,\ldots ,N\rbrace $.

To solve the optimal control problem (7), it is necessary to establish the structure of the dependence of the functional $I\left(\alpha ^{(0)},\alpha
^{(1)}\right)$ on the discrete probability distributions $\alpha
^{(0)},\alpha ^{(1)}$. However, before formulating and proving the corresponding result, we make several important remarks on the features of the constructed model and the stated optimal control problem.

\textbf{Remark 1.} The main process control scheme $\left\{\widehat{\xi}_k\right\}_{k=0}^{\infty}$ considered in this stochastic model differs from the standard control scheme adopted in classical papers \cite{SN22}, \cite{SN23} and in many subsequent studies. In this model, decisions about the choice of control are taken at some random points in time at which the process $\left\{\widehat{\xi}_k\right\}_{k=0}^{\infty}$ reaches the boundary states. In the standard control scheme of Markov and semi-Markov stochastic processes, decisions about the choice of control are taken at each moment when a change in the state of the process occurs. Thus, the problem of optimal control of the process $\left\{\widehat{\xi}_k\right\}_{k=0}^{\infty}$, which is in form the problem of tuning for a model with discrete time, is not reduced to the classical formulation of the problem of stochastic control with respect to stationary cost indicators of the form (6).

\textbf{Remark 2.} We introduce an important assumption related to the initial probability characteristics of the model. Namely, we will assume that the absorption probabilities of a Markov chain $\left\{\xi_k^{(n)}\right\}_{k=0}^{\infty}$ with an arbitrary number $n \in \{0,1,2,\dots\}$ satisfy condition $b_{l0}>0$, $b_{l1}>0$, $l\in\{2,3,\dots,N\}$; at the same time, as already noted, $b_{l0}+b_{l1}=1$, $l\in\{2,3,\dots,N\}$.This condition means that a Markov chain with an arbitrary number $n\in\{0,1,2,\dots\}$ in a finite time with a probability equal to one reaches one of its absorbing states: $\{0\}$ or $\{1\}$. For the constructed Markov model $\left\{\widehat{\xi}_k\right\}_{k=0}^{\infty}$ this condition has the following probabilistic meaning. As a result of the next external influence (control), the Markov process $\left\{\widehat{\xi}_k\right\}_{k=0}^{\infty}$ is transferred to one of the internal admissible states $\{2,3,\dots,N\}$. After this, the main process $\left\{\widehat{\xi}_k\right\}_{k=0}^{\infty}$ with a probability equal to one reaches one of its boundary states $\{0\}$ or $\{1\}$. Next, the following external influence (control) is performed, as a result of which the process will be transferred to one of the internal allowable states.
The further evolution of the process $\left\{\widehat{\xi}_k\right\}_{k=0}^{\infty}$ occurs independently of the past and according to the laws described above.
Thus, this condition ensures an infinitely long evolution of the process $\left\{\widehat{\xi}_k\right\}_{k=0}^{\infty}$ regardless of the controls made at the moments when the set of states reaches the boundary.
This property can be called the stability of the constructed stochastic model. We emphasize that it will be carried out for any strategy for controlling the process $\left\{\widehat{\xi}_k\right\}_{k=0}^{\infty}$, which is determined by probability distributions $(\alpha^{(0)},\alpha^{(0)})$.
As will be shown later, this condition also ensures the existence of a single stationary distribution of some auxiliary Markov chain embedded in the process $\left\{\widehat{\xi}_k\right\}_{k=0}^{\infty}$. In the future, we will require the fulfillment of this condition when proving the main results.

Now we can proceed to the formulation and proof of the statement about the explicit representation of the stationary cost indicator of the effectiveness of control.

\textbf{Theorem 1.}

Suppose that the following conditions are met in the stochastic model under consideration: $b_{l,0}>0,~b_{l,1}>0,~l\in\{2,3,\dots,N\}$. Then the following representation takes place for the stationary cost indicator $I=I\left(\alpha^{(0)},\alpha^{(1)}\right)$, defined by the relation (6).

$$I = I\left(\alpha ^{(0)},\alpha
^{(1)}\right)=\frac{\sum\limits_{m_{0}=2}^{N}{\sum\limits_{m_{1}=2}^{N}{A(m_{0},m_{1})\alpha
_{m_{0}}^{(0)}\alpha_{m_{1}}^{(1)}}}}{\sum\limits_{m_{0}=2}^{N}{\sum\limits_{m_{1}=2}^{N}{B(m_{0},m_{1})\alpha
_{m_{0}}^{(0)}\alpha _{m_{1}}^{(1)}}}} \eqno(8)$$
where
$$A(m_{0},m_{1})=\left[ d_{m_{0}}^{(0)}+r_{m_{0}}\right] b_{m_{1},0}
+\left[ d_{m_{1}}^{(1)}+r_{m_{1}}\right] b_{m_{0},1} \eqno(9)$$
$$B(m_{0},m_{1})=b_{m_{0},1}+b_{m_{1},0} \eqno(10)$$

Proof.

We first prove the following lemma related to the probability characteristics of the auxiliary Markov chain $\left\{\zeta_n \right\}_{n=0}^{\infty}$.

\textbf{Lemma 1.} Suppose that conditions $b_{l,0}>0,~b_{l,1}>0$, $l\in\{2,3,\dots,N\}$ are satisfied. Then the transition probability of the Markov chain $\left\{\zeta_n\right\}_{n=0}^{\infty}$ is determined by

$$
\widetilde{p}_{ij} = P\left(\zeta_{n+1}=j\vert\zeta_n=i\right)=\sum\limits_{l=2}^{N}\alpha_l^{(i)}b_{lj}~, \eqno(11)
$$
$$
i,j\in Z=\{0,1\}
$$

Proof of Lemma 1.

During the proof, various random events related to the trajectories of the main process will be considered $\left\{\widehat{\xi}_k\right\}_{k=0}^{\infty}$. In accordance with the initial assumptions about the construction of a stochastic model (Section V), all these events are defined on the probability space $\left(\Omega, \textit{A},\textbf{\textit{P}}\right)$, that is, elements of the $\sigma$-algebra $\textit{A}$.

We fix arbitrary consecutive moments of the process $\left\{\widehat{\xi}_k\right\}_{k=0}^{\infty}$ entering the boundary states: $\widehat{\nu}_n=k_n$, $\widehat{\nu}_{n+1}=k_{n+1}$ and we will consider random events associated with the trajectories of the process $\left\{\widehat{\xi}_k\right\}_{k=0}^{\infty}$ on the time interval $\left[k_n,k_{[n+1]}\right]$.

Now we fix the state $i\in Z=\{0,1\}$ and consider the random event $\left(\widehat{\xi}_{k_n}=i\right)$. Events that will be introduced later will be considered under the condition that event $\left(\widehat{\xi}_{k_n}=i\right)$ occurred, that is, on the set of elementary outcomes corresponding to this event.

Note that if condition $\left(\widehat{\xi}_{k_n}=i\right)$, is satisfied, that is, at time $k_n$ process $\left\{\widehat{\xi}_k\right\}_{k=0}^{\infty}$ is in the boundary state $i$, then by the property of the adopted stochastic model at time $k_n+1$ it will be transferred to one of the internal allowable states $l\in\{2,3,\dots,N\}$, that is, one of the system $\left(\widehat{\xi}_{k_n+1}=l\right)$, $l\in\{2,3,\dots,N\}$ will be implemented.
Thus, there is an embedding
$$
\left(\widehat{\xi}_{k_n}=i\right)\subset \bigcup\limits_{l=2}^{N}\left(\widehat{\xi}_{k_n+1}=l\right)~, i\in Z=\{0,1\}
$$

Events from system $\left\{\left(\widehat{\xi}_{k_n+1}=l\right),~l\in\{2,3,\dots,N\}\right\}$ are pairwise incompatible.
At the same time, each of them together with the event $\left(\widehat{\xi}_{k_n}=i\right)$, since as a result of the control, the transfer from the boundary state $i\in Z$ can be made to any internal state $l\in \{2,3,\dots,N\}$.

We next consider event $\left(\widehat{\xi}_{k_{n+1}}=j\right)$, $j\in Z=\{0,1\}$ - a certain boundary state. This event can be realized only when, as a result of the previous control, the main process $\left\{\widehat{\xi}_k\right\}_{k=0}^{\infty}$ was transferred to one of the internal states $l\in\{2,3,\dots,N\}$, that is, one of the system of incompatible events $\left\{\left(\widehat{\xi}_{k_n+1}=l,~l\in \{2,3,\dots,N\}\right)\right\}$ occurred.
Therefore, it can be argued that on the set of outcomes of this experiment, corresponding to the fixed condition $\left(\widehat{\xi}_{k_n}=i\right)$, there is an embedding of events
$$
\left(\widehat{\xi}_{k_{n+1}}=j\right) \subset \bigcup\limits_{l=2}^{N} \left(\widehat{\xi}_{k_n+1}=l\right),~j\in Z=\{0,1\}.
$$

From the comments made it follows that the set of total probability is applicable on the set of elementary outcomes corresponding to the event $\left(\widehat{\xi}_{k_n}=i\right)$. In this case, the role of the main event, the probability of which is necessary to determine, will play the event $\left(\widehat{\xi}_{k_{n+1}}=j\right)$, and the role of the system of incompatible hypotheses events-set $\left\{\left(\widehat{\xi}_{k_n+1}=l\right),~l\in\{2,3,\dots,N\}\right\}$. The probabilities of all specified events are determined under condition $\left(\widehat{\xi}_{k_n}=i\right)$. Based on the formula of total probability, we have
$$
P\left(\widehat{\xi}_{k_{n+1}}=j\vert\widehat{\xi}_{k_n}=i\right)=
$$
$$
=\sum\limits_{l=2}^{N}P\left(\widehat{\xi}_{k_{n+1}}=j\vert\widehat{\xi}_{k_n}=i,~\widehat{\xi}_{k_n+1}=l\right)
P\left(\widehat{\xi}_{k_n+1}=l\vert\widehat{\xi}_{k_n}=i\right)
\eqno(12)
$$

At the same time, by the property of the constructed model (see relations (2), (4) and the corresponding explanations for them).
$$
P\left(\widehat{\xi}_{k_n+1}=l\vert\widehat{\xi}_{k_n}=i\right)=\alpha_l^{(i)},~l\in\{2,3,\dots,N\},~i\in\{0,1\}.
\eqno(13)
$$

We use another property of the constructed stochastic model. After the next hitting into the boundary state and the produced control, the evolution of the process $\left\{\widehat{\xi}_k\right\}_{k=0}^{\infty}$ will depend only on the state to which the process was transferred as a result of the control. It follows that for any $i,j \in\{0,1\}$, $l\in\{2,3,\dots,N\}$ there is an equality
$$
P\left(\widehat{\xi}_{k_{n+1}}=j\vert\widehat{\xi}_{k_n}=i,~\widehat{\xi}_{k_n+1}=l\right)=
$$
$$
=P\left(\widehat{\xi}_{k_{n+1}}=j\vert\widehat{\xi}_{k_n+1}=l\right) \eqno(14)
$$

The trajectory of the process $\left\{\widehat{\xi}_k\right\}_{k=0}^{\infty}$ on the time interval $\left[k_n+1,~k_{n+1}\right]$ coincides with the trajectory of the absorbing Markov chain $\left\{\xi_k^{(n+1)}\right\}_{k=0}^{\infty}$ (see relation (5) and the corresponding explanations). Thus, the transition probability on the right-hand side of equality (14) will coincide with the absorption probability of the chain $\left\{\xi_k^{(n+1)}\right\}_{k=0}^{\infty}$.
$$
P\left(\widehat{\xi}_{k_{n+1}}=j\vert\widehat{\xi}_{k_{n+1}}=l\right)=
$$
$$
=P\left(\xi_{k_{n+1}-k_{n}-1}^{(n+1)}=j\vert\xi_{0}^{(n+1)}=l\right)=b_{lj}~, \eqno(15)
$$
$$
l\in\{2,3,\dots,N\},~j\in\{0,1\}.
$$

From (14) and (15) it immediately follows that for any fixed $i\in\{0,1\}$
$$
\textbf{\textit{P}}\left(\widehat{\xi}_{k_{n+1}}=j\vert\widehat{\xi}_{k_n}=i,~\widehat{\xi}_{k_n+1}=l\right)=b_{lj} \eqno(16)
$$
$$
l\in \{2,3,\dots,N\},~j\in \{0,1\}~.
$$

Substituting (13) and (16) into relation (12), we obtain
$$
\textbf{\textit{P}}\left(\widehat{\xi}_{k_{n+1}}=j\vert\widehat{\xi}_{k_n}=i\right)=
$$
$$
=\sum\limits_{l=2}^{N} \alpha_l^{(i)}b_{lj},~i,j\in\{0,1\}~. \eqno(17)
$$

It remains to be noted that, within the framework of the adopted model, there is a coincidence of events $\left(\widehat{\xi}_{k_n}=s\right)=\left(\zeta_n=s\right)$, $s\in\{0,1\}, n=0,1,2,\dots$.
The assertion of Lemma 1 follows from equality (17).

Proof of Theorem 1.

Consider the properties of the auxiliary Markov chain $\left\{\zeta_n\right\}_{n=0}^{\infty}$. If conditions $b_{l0}>0,~b_{l1}>0$, $l\in\{2,3,\dots,N\}$, are satisfied, then the statement of Lemma 1 (relation (11)) implies that $\widetilde{p}_{ij}>0,~i,j\in\{0,1\}$. Thus, the states of a given chain are interconnected and form one class, that is, this Markov chain is irreducible. Moreover, the set of states of the chain $Z=\{0,1\}$ is finite; it follows from this that all states are recurrent and positive.

As is well known (\cite{SN16}, \cite{SN17}), a Markov chain with the indicated properties has a unique stationary distribution $\pi=(\pi_0,\pi_1)$, which satisfies the system of equations
$$
\pi_0 = \pi_0 \widetilde{p}_{00} + \pi_1 \widetilde{p}_{10}
$$
$$
\pi_1 = \pi_0 \widetilde{p}_{01} + \pi_1 \widetilde{p}_{11}
$$
$$
\pi_0+\pi_1=1
\eqno(18)
$$

The solution of the system of equations (18) is
$$
\pi_0=\dfrac{\widetilde{p}_{10}}{1-\widetilde{p}_{00}+\widetilde{p}_{10}}=\dfrac{\widetilde{p}_{10}}{\widetilde{p}_{01}+\widetilde{p}_{10}},
$$
$$
\pi_1=\dfrac{1-\widetilde{p}_{00}}{1-\widetilde{p}_{00}+\widetilde{p}_{10}}=\dfrac{\widetilde{p}_{01}}{\widetilde{p}_{01}+\widetilde{p}_{10}}
\eqno(19)
$$

Substituting into equation (19) the formulas for transition probabilities (11) and we obtain
$$
\pi_0 = \dfrac{\sum\limits_{l=2}^{N}\alpha_l^{(1)}b_{l0}}{\sum\limits_{l=2}^{N}\alpha_l^{(0)}b_{l1}+\sum\limits_{l=2}^{N}\alpha_l^{(1)}b_{l0}}
\eqno(20)
$$
$$
\pi_1 = \dfrac{\sum\limits_{l=2}^{N}\alpha_l^{(0)}b_{l1}}{\sum\limits_{l=2}^{N}\alpha_l^{(0)}b_{l1}+\sum\limits_{l=2}^{N}\alpha_l^{(1)}b_{l0}} \eqno(21)
$$

We now define the cost additive functional associated with the auxiliary Markov chain $\left\{\zeta_n\right\}_{n=0}^{\infty}$. We will assume that the income $\rho_i$, obtained by a single stay of this process in state $i\in Z=\{0,1\}$, coincides with the average profit generated in the system under consideration for the time elapsed from the moment the main stochastic process $\left\{\widehat{\xi}_k\right\}$ hits the boundary state $i$ until the next state falls into the boundary state. Let $\Delta v_n$ be the random value of the profit generated in the system for the specified period of time. Then
$$
\rho_i = E\left[\Delta v_n\vert\widehat{\xi}_{k_n}=\zeta_n=i\right],~i\in Z = \{0,1\} \eqno(22).
$$

We calculate the magnitude of the conditional expectation, which is located on the right-hand side of equality (22).
By the property of mathematical expectation
$$
\rho_i = E\left[\Delta v_n\vert\widehat{\xi}_{k_n}=\zeta_n=i\right]=
$$
$$
\sum\limits_{l=2}^{N} E\left[\Delta v_n\vert\widehat{\xi}_{k_n}=\zeta_n=i,~\widehat{\xi}_{k_n+1}=l\right]\times
$$
$$
\times P\left(\widehat{\xi}_{k_n+1}=l\vert\widehat{\xi}_{k_n}=\zeta_n=i\right),
$$
$$
i\in\{0,1\} \eqno(23)
$$

For the stochastic model under consideration, the increment in profit over the time between successive hits of the main process $\left\{\widehat{\xi}_k\right\}_{k=0}^{\infty}$ into boundary states is made up of the costs associated with transferring this process to one of the internal states and random income received during the free evolution until the next hit to one of the boundary states.

From here follows
$$
E\left[\Delta v_n\vert\widehat{\xi}_{k_n}=\zeta_n=i,~\widehat{\xi}_{k_n+1}=l\right] = d_{l}^{(i)}+r_l,
$$
$$
l\in \{2,3,\dots,N\},~i\in\{0,1\}
\eqno(24)
$$

From (23), taking into account (24) and (13), we obtain
$$
rho_i = E\left[\Delta v_n\vert\widehat{\xi}_{k_n}=\zeta_n=i\right] = \sum\limits_{l=2}^{N} \left[d_l^{(i)}+r_l\right]\alpha_l^{(i)},~i\in\{0,1\}
\eqno(25)
$$

As already noted, the auxiliary Markov chain $\left\{\zeta_n\right\}_{n=0}^{\infty}$ is irreducible, recurrent, and positive.
Then the above ergodic theorem for the additive cost functional associated with the stochastic model under consideration and the Markov chain $\left\{\zeta_n\right\}_{n=0}^{\infty}$ is valid. From the statement of this theorem and relation (6) it follows
$$
I = \rho_0 \pi_0 + \rho_1 \pi_1
\eqno(26)
$$

Substituting into (26) formulas (20), (21), (25), we obtain the following representation
$$
I =
$$
\small $$
\dfrac{
\sum\limits_{m=2}^{N}\alpha_m^{(0)}\left[d_m^{(0)}+r_m\right]\sum\limits_{l=2}^{N}\alpha_l^{(1)}b_{l0}+
\sum\limits_{m=2}^{N}\alpha_m^{(1)}\left[d_m^{(1)}+r_m\right]\sum\limits_{l=2}^{N}\alpha_l^{(0)}b_{l1}}
{\sum\limits_{l=2}^{N}\alpha_l^{(0)}b_{l1}+\sum\limits_{l=2}^{N}\alpha_l^{(1)}b_{l0}}
\eqno(27)
$$ \normalsize

Let us transform the expressions in the numerator and denominator of the right side of formula (27). Consider the numerator
$$
I_1\left(\alpha^{(0)},\alpha^{(1)}\right)=
$$
\small $$\sum\limits_{m=2}^{N}\alpha_m^{(0)}\left[d_m^{(0)}+r_m\right]\sum\limits_{l=2}^{N}\alpha_l^{(1)}b_{l0}+
\sum\limits_{m=2}^{N}\alpha_m^{(1)}\left[d_m^{(1)}+r_m\right]\sum\limits_{l=2}^{N}\alpha_l^{(0)}b_{l1}
$$ \normalsize

We write the products of sums in the form of a double sum of pairwise products
$$
I_1\left(\alpha^{(0)},\alpha^{(1)}\right) = \sum\limits_{m=2}^{N}\sum\limits_{l=2}^{N}\left[d_m^{(0)}+r_m\right]b_{l0}\alpha_m^{(0)}\alpha_l^{(1)}+
$$
$$
+\sum\limits_{m=2}^{N}\sum\limits_{l=2}^{N}\left[d_m^{(1)}+r_m\right]b_{l1}\alpha_l^{(0)}\alpha_m^{(1)} \eqno(28)
$$

Now we carry out redesignations of the summation indices in the first and second terms of the right-hand side of equality (28). In the first double sum we put $m=m_0,~l=m_1$, and in the second $l=m_0,~m=m_1$. After that, combine both amounts into one and get
$$
I_1\left(\alpha^{(0)},\alpha^{(1)}\right) =
$$
\small $$=\sum\limits_{m_0=2}^{N}\sum\limits_{m_1=2}^{N} \left[\left[d_{m_0}^{(0)}+r_{m_0}\right]b_{m_1,0}+\left[d_{m_1}^{(1)}+r_{m_1}\right]b_{m_0,1}\right]\alpha_{m_0}^{(0)}\alpha_{m_1}^{(1)}
\eqno(29)
$$ \normalsize

Now consider the representation of the denominator on the right side of formula (27)
$$
I_0\left(\alpha^{(0)},\alpha^{(1)}\right) =
\sum\limits_{l=2}^{N}\alpha_l^{(0)}b_{l1} + \sum\limits_{l=2}^{N}\alpha_l^{(1)}b_{l0}
$$

We perform the identity transformation of the expression $I_0\left(\alpha^{(0)},\alpha^{(1)}\right)$ taking into account the normalization conditions for probability distributions $\alpha^{(0)},\alpha^{(1)}$.
\small $$
I_0\left(\alpha^{(0)},\alpha^{(1)}\right) = \sum\limits_{m=2}^{N} \alpha_m^{(1)} \sum\limits_{l=2}^{N} \alpha_l^{(0)}b_{l1} + \sum\limits_{m=2}^{N} \alpha_m^{(0)} \sum\limits_{l=2}^{N} \alpha_l^{(1)}b_{l0} =
$$ \normalsize
$$
= \sum\limits_{m=2}^{N} \sum\limits_{l=2}^{N} b_{l1} \alpha_l^{(0)}\alpha_m^{(1)} + \sum\limits_{m=2}^{N} \sum\limits_{l=2}^{N} b_{l0} \alpha_m^{(0)}\alpha_l^{(1)} \eqno(30)
$$

Next, we will redesign the summation indices in the first and second terms of the right-hand side of equality (30) in the same way as was done above when converting the numerator, and then combine these sums into one. Then
$$
I_0\left(\alpha^{(0)},\alpha^{(1)}\right) = \sum\limits_{m_0=2}^{N} \sum\limits_{m_1=2}^{N} \left[b_{m_0,1}+b_{m_1,0}\right]\alpha_{m_0}^{(0)}\alpha_{m_1}^{(1)}
\eqno(31)
$$

From equality (27), taking into account the transformations of expressions for the numerator and denominator, given by formulas (29), (31), we obtain the representation for the stationary cost indicator $I\left(\alpha^{(0)}, \alpha^{(1)}\right)$ in the form (8). In this case, the functions $A(m_0, m_1)$, $B(m_0, m_1)$  are determined by formulas (9), (10), respectively. Theorem 1 is proved.

From the statement of Theorem 1, it follows that in the stochastic model under consideration, the stationary cost indicator of control efficiency $I\left(\alpha^{(0)}, \alpha^{(1)}\right)$ is represented as a linear fractional integral functional defined on a set of pairs of discrete probability distributions $\left(\alpha^{(0)}, \alpha^{(1)}\right)$, each of which defines a control strategy. In this regard, to solve the optimal control problem, which is formulated as an extremal problem (7), it is necessary to use the theoretical results for the unconditional extremum problem of a functional of this form.  summary of these results will be given in the next section.

\section{Extreme problem for fractional-linear integral functionals defined on a set of sets of discrete probability distributions}

The solution of the general problem of unconditional extremum of a linear fractional integral functional defined on a set of probability measures was described in the work of P.V. Shnurkov \cite{SN26}. Subsequently, the proof of the corresponding result was published in \cite{SN27}. To solve the extremal problem (7) arising in the course of this study, we will need a special version of the general statement formulated in \cite{SN26}, \cite{SN27}. In this variant, the objective functional of the posed extremal problem is defined on the set of finite sets of discrete probability distributions. Then the multidimensional integrals in the numerator and denominator of the target functional are converted into multidimensional sums. Since such a functional is a special version of the general fractional-linear integral functional, we will use the same name for it with the addition of the word "discrete".In this section, we present the formulation of an extremal problem for a fractional-linear integral discrete functional and a theorem on solving this problem. The statement of this theorem will constitute the theoretical basis for solving the extremal problem (7).

We adopt the following notational convention. Designations for various mathematical objects introduced in Section VIII will be used only within the framework of this section. The connection of these objects with the objects of the stochastic tuning problem considered in this study will be established in the next part of the paper.

We consider a set of discrete sets $U_i={1,2,\dots,n_i}$, $i=1,2,\dots,N$, $N<\infty$.
The number of elements in each set~$U_i$ can be either finite or countable:
$n_i\leq\infty$, $i=1,2,\dots,N$.
In what follows, these sets are interpreted as sets of admissible solutions (controls)
accepted in different states of the stochastic model,
but in this section, they are abstract.
On each set $U_i$, we introduce a collection
of all possible probability distributions of the form
$\alpha^{(i)}=(\alpha_1^{(i)},\alpha_2^{(i)},\dots,\alpha_{n_i}^{(i)})$,
$\alpha_s^{(i)}\geq 0$, $s=1,2,\dots,n_i$;
$\sum\limits_{s=1}^{n_i}\alpha_s^{(i)}=1$, $i=1,2,\dots,N$.
We denote such a set of probability distributions defined on~$U_i$
by~$\Gamma_{d}^{(i)}$, $i=1,2,\dots,N$.
Further, we consider the Cartesian product of spaces
$U=U_1\times U_2\times\dots\times U_N$.
Following the classical scheme of introducing the measure
on a Cartesian product of spaces \cite{SN28},
we introduce the probability measure on~$U$
as the product of probability measures on the spaces
$U_1,U_2,\dots,U_N$ determined by the distributions
$\alpha^{(1)},\alpha^{(2)},\dots,\alpha^{(N)}$.
Thus, the probability measure on~$U$ is given
by the set of probability distributions
$\alpha^{(1)}$, $\alpha^{(2)},\dots,\alpha^{(N)}$.
We denote the set of probability measures on~$U$ by $\Gamma_{d}$.
Several additional conditions related to the set~$\Gamma_{d}$
are described below in the statement of the extremum problem.

Similarly to \cite{SN26}, we introduce the notion of degenerate discrete probability distribution.

\textbf{Definition 2.}
A probability distribution $\alpha^{(i)*}(k_i)$
is said to be \textit{degenerate} if
$\alpha_{k_i}^{(i)}=1$, $\alpha_l^{(i)}=0$, $l=1,2,\dots,n_i$, $l\neq k_i$.
A point $k_i\in U_i$ is called a \textit{point of concentration}
of a degenerate distribution $\alpha^{(i)*}(k_i)$.
As is known, a degenerate distribution corresponds to
the deterministic quantity taking the value~$k_i$.

We denote the set of degenerate probability distributions defined on~$U_i$ by
$\Gamma_{d}^{(i)*}$, $i=1,2,\dots,N$.
Obviously, there is a one-to-one correspondence between the sets
$\Gamma_d^{(i)*}$ and $U_i$, $i=1,2,\dots,N$.
We accordingly denote
the set of all degenerate probability measures defined on the set~$U$
by~$\Gamma_{d}^{*}$.
Each degenerate measure in the set $\Gamma_{d}^{*}$
is defined by a set of degenerate distributions
$\alpha^{(1)*},\alpha^{(2)*},\dots,\alpha^{(N)*}$.

We assume that two numerical functions are defined on the set~$U$:
$$
A(k_1,k_2,\dots,k_N): U\rightarrow R,\quad
$$
$$
B(k_1,k_2,\dots,k_N): U\rightarrow R,\quad
$$
$$
\text{where $k_i\in U_i$, $i=1,2,\dots,N$}.
$$

We note that the integral over a discrete measure defined on a discrete set
can be transformed into a sum,
and the corresponding multidimensional integral over the measure
generated by the product of initial measures
defined on the Cartesian product of discrete spaces
becomes a multidimensional sum.
According to this, we introduce the following definition by analogy with \cite{SN26}.

\textbf{Definition 3.}

A \textit{linear-fractional integral functional} (\textit{in the discrete version})
or simply a \textit{discrete linear-fractional integral functional}
defined on a set of collections of discrete probability distributions $\Gamma_{d}$
is defined to be the mapping
$I(\alpha^{(1)},\alpha^{(2)},\dots,\alpha^{(N)}): \Gamma_{d}\rightarrow R$
given by the expression
$$
I(\alpha^{(1)},\alpha^{(2)},\dots,\alpha^{(N)})=
$$
$$
=\dfrac{\sum\limits_{k_1=1}^{n_1}\sum\limits_{k_2=1}^{n_2}\dots\sum\limits_{k_{N}=1}^{n_N}
A(k_1,k_2,\dots,k_{N})\alpha_{k_1}^{(1)}\alpha_{k_2}^{(2)}\dots\alpha_{k_{N}}^{(N)}}
{\sum\limits_{k_1=1}^{n_1}\sum\limits_{k_2=1}^{n_2}\dots\sum\limits_{k_{N}=1}^{n_N}
B(k_1,k_2,\dots,k_{N})\alpha_{k_1}^{(1)}\alpha_{k_2}^{(2)}\dots\alpha_{k_{N}}^{(N)}} \eqno(32)
$$

\textbf{Definition 4.}

A function $C(k_1,k_2,\dots,k_N):U\rightarrow R$ defined by the expression
$$
C(k_1,k_2,\dots,k_N)=\dfrac{A(k_1,k_2,\dots,k_N)}{B(k_1,k_2,\dots,k_N)}~, \eqno(33)
$$
is the \textit{test} function of the discrete linear-fractional integral functional
$I(\alpha^{(1)},\alpha^{(2)},\dots,\alpha^{(N)})$ given by formula~(32).

Let us formulate the corresponding extremum problem for
$I(\alpha^{(1)},\alpha^{(2)},\dots,\alpha^{(N)})$ of the form (32)
on a set of collections of discrete probability distributions $\Gamma_d$:
\small $$
I(\alpha^{(1)},\alpha^{(2)},...,\alpha^{(N)}) \to extr,~~ (\alpha^{(1)},\alpha^{(2)},...,\alpha^{(N)})\in \Gamma_d
\eqno{(34)}
$$ \normalsize

We assume that some preliminary conditions similar to the corresponding conditions
introduced when solving the extremum problem
for a linear-fractional integral functional of general structure considered in \cite{SN26}, \cite{SN27}
are satisfied for the above-posed extremum problem~(34).
Let us specify these conditions.
\begin{enumerate}
\item
The functionals of discrete probability distributions,
which determine the numerator and denominator in expression~(32),
are defined for any probability distributions
$(\alpha^{(1)},\alpha^{(2)},\dots,\alpha^{(N)})\in \Gamma_d$ as
\begin{equation*}
\begin{aligned}
I_1(\alpha^{(1)},\alpha^{(2)},...,\alpha^{(N)})=\\
=\sum\limits_{k_1=1}^{n_1}\sum\limits_{k_2=1}^{n_2}...\sum\limits_{k_{N}=1}^{n_N}
A(k_1,k_2,...,k_{N})\alpha_{k_1}^{(1)}\alpha_{k_2}^{(2)}...\alpha_{k_{N}}^{(N)}
\end{aligned}
\eqno{(35)}
\end{equation*}
\begin{equation*}
\begin{aligned}
I_2(\alpha^{(1)},\alpha^{(2)},...,\alpha^{(N)})=\\
=\sum\limits_{k_1=1}^{n_1}\sum\limits_{k_2=1}^{n_2}...\sum\limits_{k_{N}=1}^{n_N}
B(k_1,k_2,...,k_{N})\alpha_{k_1}^{(1)}\alpha_{k_2}^{(2)}...\alpha_{k_{N}}^{(N)}
\end{aligned}
\eqno{(36)}
\end{equation*}
In other words, the numerical series in the right-hand sides of expressions~(35) and~(36)
are assumed to converge.
\item
For any discrete probability distributions
$(\alpha^{(1)},\alpha^{(2)},\dots,\alpha^{(N)})\in\Gamma_d$,
the functional $I_2(\alpha^{(1)},\alpha^{(2)},\dots,\alpha^{(N)})$
does not vanish, namely,
$$
I_2(\alpha^{(1)},\alpha^{(2)},\dots,\alpha^{(N)})=
\sum\limits_{k_1=1}^{n_1}\sum\limits_{k_2=1}^{n_2}\dots\sum\limits_{k_{N}=1}^{n_N}
$$
$$
B(k_1,k_2,\dots,k_{N})\alpha_{k_1}^{(1)}\alpha_{k_2}^{(2)}\dots\alpha_{k_{N}}^{(N)}\ne
$$
$$
\ne  0.
$$
\item
The set of collections of degenerate probability distributions $\Gamma_d^*$
is completely contained in the set $\Gamma_d:\Gamma_d^*\subset\Gamma_d$.
\end{enumerate}

\textbf{Remark 3.}
As in the general version~[26],  conditions~2 and~3 imply that
the function $B(k_1,k_2,\dots,k_N)$ is strictly of constant sign
for all $(k_1,k_2,\dots,k_N)\in U$.
At the same time,
if this condition related to the character of the function $B(k_1,k_2,\dots,k_N)$
is satisfied, then condition~2 is satisfied automatically.
In \cite{SN27}, it is specially noted that the condition
of being strictly of constant sign
(and, in particular, of being strictly positive)
for the function $B(k_1,k_2,\dots,k_N)$ is natural for
optimal control problems for stochastic processes.
In this connection, it is required that this condition is satisfied
in the fundamental theorem on the solution of extremum problem~(34).

\textbf{Definition 4.}
The set of collections of discrete probability distributions $\Gamma_{d}$
is said to be \textit{admissible} in extremum problem~(34)
if conditions~1 and~3 in the system of preliminary conditions
are satisfied.

We now formulate the fundamental theorem on the solution of extremum problem~(34)
which is a particular case of Theorem~1 formulated in \cite{SN26}.
We restrict our consideration to the first assertion in this theorem
as the most important for solving the optimal control problem
considered in the present paper.

\textbf{Theorem 2.}

Assume that the set of collections of discrete probability distributions
$\Gamma_{d}$ in extremum problem~{\rm(34)} is admissible
and the function $B(k_1,k_2,\dots,k_N)$
in discrete linear-fractional integral functional~{\rm(32)}
is strictly of constant sign
{\rm(}strictly positive or strictly negative{\rm)}
for all values of the arguments $(k_1,k_2,\dots,k_N)\in U$.
Assume also that the test function
of the discrete linear-fractional integral functional
$C(k_1,k_2,\dots,k_{N})=\dfrac{A(k_1,k_2,\dots,k_{N})}{B(k_1,k_2,\dots,k_{N})}$
attains a global extremum {\rm(}maximum or minimum{\rm)}
on the set~$U$ at a fixed point $(k_1^*,k_2^*,\dots,k_{N}^*)$.
Then the solution of the corresponding extremum problem~{\rm(34)}
for the maximum or minimum exists
and is attained on the set of degenerate probability distributions
$(\alpha^{(1)^*},\alpha^{(2)^*},\dots,\alpha^{(N)^*})$
concentrated at the respective points $k_1^*,k_2^*,\dots,k_{N}^*$
and the following relations are satisfied:
\begin{equation*}
\begin{aligned}
\max\limits_{(\alpha^{(1)},\alpha^{(2)},...,\alpha^{(N)})\in\Gamma_d}
I(\alpha^{(1)},\alpha^{(2)},...,\alpha^{(N)})=\\
=\max\limits_{(\alpha^{(1)^*},\alpha^{(2)^*},...,\alpha^{(N)^*})\in\Gamma_d^*}
I(\alpha^{(1)^*},\alpha^{(2)^*},...,\alpha^{(N)^*})= \\
=\max\limits_{(k_1,k_2,...,k_{N})\in U}
\dfrac{A(k_1,k_2,...,k_{N})}{B(k_1,k_2,...,k_{N})}=\\
=\dfrac{A(k_1^*,k_2^*,...,k_{N}^*)}{B(k_1^*,k_2^*,...,k_{N}^*)},
\end{aligned}
\eqno{(37)}
\end{equation*}
if the global maximum of the test function is attained
at the point $(k_1^*,k_2^*,\dots,k_{N}^*)$;
\begin{equation*}
\begin{aligned}
\min\limits_{(\alpha^{(1)},\alpha^{(2)},...,\alpha^{(N)})\in\Gamma_d}
I(\alpha^{(1)},\alpha^{(2)},...,\alpha^{(N)})=\\
=\min\limits_{(\alpha^{(1)^*},\alpha^{(2)^*},...,\alpha^{(N)^*})\in\Gamma_d^*}
I(\alpha^{(1)^*},\alpha^{(2)^*},...,\alpha^{(N)^*})= \\
=\min\limits_{(k_1,k_2,...,k_{N})\in U}
\dfrac{A(k_1,k_2,...,k_{N})}{B(k_1,k_2,...,k_{N})}=\\
=\dfrac{A(k_1^*,k_2^*,...,k_{N}^*)}{B(k_1^*,k_2^*,...,k_{N}^*)},
\end{aligned}
\eqno{(38)}
\end{equation*}
if the global minimum of the test function is attained
at the point $(k_1^*,k_2^*,\dots,k_{N}^*)$.

\section{Solving the problem of optimal control in a stochastic Markov model with discrete time}

Let us turn to solving the extremal problem (7) for the objective functional $I\left(\alpha^{(0)}, \alpha^{(1)}\right)$, defined by formula (8). In its analytic form, the functional $I\left(\alpha^{(0)}, \alpha^{(1)}\right)$ belongs to the form of fractional-linear integral discrete functionals defined by relation (32). The roles of the spaces of admissible solutions (controls) are played by finite sets $U_0=U_1=\{2,3,\dots,N\}$ and their Cartesian product $U=U_0\times U_1$, which is a set of pairs: $U=\left\{(m_0,m_1):~m_0\in U_0,~m_1\in U_1\right\}$. The discrete probability distributions $\alpha^{(0)}=\left(\alpha_l^{(0)},~l=2,3,\dots,N\right)$, $\alpha^{(1)}=\left(\alpha_l^{(1)},~l=2,3,\dots,N\right)$ are defined on sets $U_0, U_1$ respectively. The main function of a linear fractional integral discrete functional $I\left(\alpha^{(0)}, \alpha^{(1)}\right)$  is determined by the formula (see general formula (33)):
$$
C(m_{0},m_{1})=\frac{A(m_{0},m_{1})}{B(m_{0},m_{1})}~, \eqno(39)
$$
where the functions $A(m_{0},m_{1})$, $B(m_{0},m_{1})$ are given by equalities (9), (10) respectively.

Note that in the problem under consideration, function $C(m_{0},m_{1})$  is defined on a finite set of argument values $U$.
Thus, this function achieves its minimum and maximum values (global extrema) on the set $
U$.
Recall also that in the stochastic model under consideration, the probability characteristics $b_{m_0,1},~b_{m_1,0}$ were assumed to be strictly positive for all values of $m_0\in U_0 = \{2,3,\dots,N\}$, $m_1\in U_1 = \{2,3,\dots,N\}$ (see note 2). It follows that the condition is satisfied.
$$
B(m_{0},m_{1})=b_{m_{0},1}+b_{m_{1},0} >0, (m_{0},m_{1})\in U
$$
Thus, all the conditions of the theorem on the extremum of a linear-fractional integral functional defined on the set of discrete probability distributions are fulfilled (Theorem 2). The solution of the extremal problem (7) exists (separately both for the maximum problem and for the minimum problem) and is achieved on degenerate discrete distributions concentrated at the points $m_{0}^{*},m_{1}^{*}$.
In this case, $(m_{0}^{*},m_{1}^{*})$ is the point at which the corresponding global extremum of the main function $C(m_{0},m_{1})$ , determined by formulas (39), (9) and (10).

Thus, the solution of the optimal control problem in this stochastic model is a pair of deterministic values of the control parameters $(m_{0}^{*},m_{1}^{*})$, delivering the extremum of the explicitly specified function $C(m_{0},m_{1})$. The problem of optimal control has been completely solved.

\section{Conclusion. The significance of the results and their possible applications}

Let's make some general remarks about the significance of the constructed mathematical model and the application of the results obtained. Research of the phenomenon of intervention in the economy is one of the important problems of applied mathematics, which is confirmed by the numerous studies mentioned in the second section. In the present paper, the mathematical model of intervention as a stochastic process with control was first constructed. This model allows describing intervention as an external influence on the economic system (grain market or foreign exchange market). Mathematically, this influence is expressed by two discrete probability distributions. Their elements represent the likelihood of transferring the main process (the price of grain or the price of the currency in the relevant free market) from any marginal inadmissible state to one of the internal admissible states.

The theoretical control problem can be interpreted as the problem of finding such probability distributions that deliver a maximum to some efficiency indicator. In terms of its economic meaning, this indicator represents the average specific profit generated in this economic system when it evolves in a stable (stationary) regime.

As a result of the theoretical investigation, it is established that the optimal probability distributions describing the external influence (control) are degenerate distributions. Since degenerate distributions correspond to deterministic values, this means that controls should be determined deterministically. In other words, interventions must be conducted in such a way that when the lower or upper unacceptable levels are reached (in the accepted notations, these are states 0 and 1, respectively), the main process is transferred to some fixed state or to one of the fixed admissible levels.

We should especially note that the optimal deterministic values of controls or levels of states of the main process, into which it should be translated as a result of the intervention, can be defined as the point of achieving the global maximum of some function of two integer variables that takes a finite number of values. For this function, an explicit analytic representation is obtained through the initial characteristics of the model.

From the point of view of practical applications, the results obtained open a new approach to the optimal organization of interventions in grain and currency markets. Having determined the initial characteristics of the model described in Section VI, it is possible to calculate in advance the optimal volumes of the interventions that are conducted, under which the main controlled process is transferred to one of the given levels.

It should be noted that the theoretical results obtained can be used to solve the problem of optimal control or regulation of technical systems. In many technical and, in particular, electronic systems, the change of the basic parameter in time can be described as a stochastic process. As such a process, the Markov chain is used in this model. To ensure the performance of the system functions, it is necessary to maintain the specified main parameter within the specified limits. Thus, the corresponding stochastic process must be in a given subset of the set of states, which can be called admissible in this model. If this process reaches one of the boundaries of the admissible subset, that is, one of the unacceptable states, it is necessary to perform some control action or system configuration, as a result of which the process (or the main parameter) must be returned to one of the states of the admissible subset. Mathematically, the control procedure can be described by two discrete probability distributions, which determine the translations of the main process from some boundary into one of the internal admissible states. It is natural to put the mathematical problem of optimal regulation or optimal adjustment of this technical system as the task of finding a pair of discrete probability distributions delivering an extremum to a stationary cost indicator of the system's performance. Thus, the problem of optimal tuning of the initial technical system formally represents the same problem of optimal stochastic control, which is the theoretical basis for solving the problem of optimizing interventions in economic systems.

To sum up, in the conducted research we developed and analyzed a universal probabilistic model that can be used to describe various systems in the economy and in technology. A theoretical solution of the optimal stochastic control problem arising in this case is obtained. The results of the research can serve as a basis for solving the corresponding problems of optimal control in various models related to interventions in the economy and optimal adjustment of parameters in technical systems.

\end{document}